\documentclass[aps,twocolumn,prb,amsmath,amssymb,superscriptaddress,longbibliography]{revtex4-2}
\usepackage{graphicx}

\usepackage{amsmath}
\usepackage{amssymb}
\usepackage{amsthm}
\usepackage{amsfonts}
\usepackage{enumerate}

\usepackage{centernot}
\usepackage{tikz}
\usepackage[colorlinks=true,citecolor=blue,linkcolor=blue]{hyperref}
\usepackage{subfigure}

\definecolor{LL-color}{named}{magenta}

\newcommand{\affiliationSICNU}{Department of Physics, Institute of Solid State Physics and Center for Computational Sciences, Sichuan Normal University, Chengdu, Sichuan 610066, China}

\newcommand{\affiliationFDU}{Department of Physics, Fudan University, Shanghai 200433, China}

\newcommand{\affiliationUSST}{College of Science, University of Shanghai for Science and Technology, Shanghai 200093, China}

\newcommand{\affiationITP}{Institute of Theoretical Physics, Chinese Academy of Science, Beijing 100080, China}

\begin{document}

\title{Consistent gauge theories for the slave particle representation of the strongly correlated $t$-$J$ model
}
 
 \author{Xi Luo}
 \thanks{These two authors contribute equally}
 \affiliation {\affiliationUSST}
 \author{Tao Shi}
 \thanks{These two authors contribute equally}
  \affiliation{\affiationITP}
 \author{Yue Yu}
 \thanks{Correspondence to: yuyue@fudan.edu.cn}
 \affiliation{\affiliationFDU}
\author{Long Liang}
\thanks{Correspondence to: longliang@sicnu.edu.cn}
\affiliation{\affiliationSICNU}

 \date{\today}
 
 \begin{abstract}
 	{We  aim to clarify the confusion and inconsistency in our recent works \cite{LLY,LYL}, and to address the
incompleteness therein.  In order to avoid the {ill-defined nature} of the free propagator of the gauge field in the ordered states of the $t$-$J$ model, we adopted  a gauge fixing {that was not of the} Becchi-Rouet-Stora-Tyutin (BRST) exact form in {our} previous work \cite{LYL}.	 This {led} to the {situation where} Dirac's second-class constraints, {namely,} the slave particle number constraint and the Ioffe-Larkin current constraint, {were} not rigorously obeyed. 	 Here we show that a consistent gauge fixing condition {that enforces} the exact constraints   {is BRST-exact in our theory}.	 An example is the Lorenz gauge. On the other hand, we prove that although the free propagator of the gauge field in the Lorenz gauge 	   is ill-defined, the full propagator is still well-defined. This implies that the strongly correlated $t$-$J$ model {can be} exactly mapped to a perturbatively controllable theory 		{within} the slave particle representation.   }
        
 \end{abstract}
 
\maketitle

\section{Introduction}

In our previous works~\cite{LLY,LYL}, we solved the two-dimensional strongly correlated problem of the electron system by 
the slave particle representation. A strongly correlated electron system is exactly mapped to a weak coupling 
slave particle system by exactly dealing with the local constraints which are either  Dirac's first-class ones when
 the system is {in} the atomic limit of the electrons~\cite{LLY} or the second-class ones in the mean field states~\cite{LYL}. 
 The essential step {for} the exact mapping  is taking a gauge fixing condition that is consistent with the local 
 constraints.  For the first-class constraint, the consistent gauge theory was {established} by Fradkin, Vilkovisky, 
 and Batalin (FVB)~\cite{FV1,FV2,BV} half a century ago. We explained their theory {in} a comprehensible language {for}
 the condensed matter physicists~\cite{LLY,LYL}.  The gauge fixing term {added} to the Lagrangian of the gauge theory is {in} a Becchi-Rouet-Stora-Tyutin (BRST)~\cite{BRS,Ty,Ty1} exact form, while the BRST charge {acting on} the 
 physical state  exactly {enforces} both the local constraints and the gauge fixing condition~\cite{Weinberg}. 
  The first-class constraint's problem has been completely solved so that we will not {concern ourselves with it} 
  in this {paper} anymore.  

There is no a systematic way to consistently solve  a gauge theory with the second-class constraints. We have solved the mean field theory where the current constraint is the second-class one in the $t$-$J$ model by considering the BRST symmetry \cite{LYL}. In principle, we have obtained a consistent gauge theory with the second-class constraint. We found that the Lorenz gauge is a consistent one while the other familiar ones, {such as} the axial gauge or the Coulomb gauge, are not.
   However, {when performing} perturbation calculations, we encountered a problem: the free {propagator} of the gauge field is ill-defined because {of} the skew $U(1)$ gauge symmetry of the Lorenz gauge fixing term. We modified the gauge fixing term so that the ill-defined problem of the free propagator {was resolved} while it {remained} BRST invariant. We then {performed} the perturbation calculation for the strange metal phase of the $t$-$J$ model and obtained some results that are consistent with the experimental measurements {on} cuprates. 
However, we did not check {whether} the gauge fixing term we used {was} a BRST exact form or not. 
{If this BRST invariant term is BRST-closed but not an exact form, the added term  {may not be} in the same BRST cohomology 
class {as} the vanishing pure gauge field Lagrangian $L_0(a_\mu)=0$ before the gauge fixing.} {For example}, we can add a Maxwell term plus the Lorenz gauge fixing term and the ghost term to $L_0(a_\mu)=0$, 
which is BRST closed but not exact. This obviously {introduces} additional dynamics to the system and changes the physics.
   On the other hand, although the free propagator of the gauge field under the Lorenz gauge is ill-defined, 
   can we still {perform} the perturbation calculation in this case? In this {paper}, we will clarify  the confusion and 
    inconsistency in the previous work and to {address} the incompleteness therein. We attach an appendix where we show the BRST formalism of the $t$-$J$ model on a lattice rather than the continuum limit  {(see Appendix \ref{app1})}.

	\section{Brief Review on the mean field theory of the $t$-$J$ model}
	
	 We consider the $t$-$J$ model 
	with the Hamiltonian on a square lattice,
\begin{eqnarray}
  H_{t-J}=-t\sum_{\langle ij\rangle,\sigma}c^\dag_{i\sigma}c_{j\sigma}+J\sum_{\langle ij\rangle}({\boldsymbol S}_i\cdot{ {\boldsymbol{S}}}_j-\frac{1}4 n_in_j), \label{tJH}
\end{eqnarray}
where $c_{i\sigma}$ is the electron annihilation operator at a lattice site $i$ with spin $\sigma$;  $S^a_i=\frac{1}2 \sum_{\sigma,\sigma'}c^\dag_{i\sigma}\sigma^a_{\sigma\sigma'}c_{i\sigma'}$ are the spin operators, and $\sigma^a$ ($a=x,y,z$) are Pauli matrices.  The hopping amplitude $t$ and the exchange amplitude $J$ are fixed {for} the nearest neighbor sites.  The constraint is that there is no double occupation at each lattice site, i.e., $c^\dag_ic_i\leq 1$ for all $i$ with a fixed total electron number.  
	
In the slave boson representation,  the electron operator is decomposed into the fermionic spinon and bosonic holon, $c^\dag_{i\sigma}= f^\dag_{i\sigma}h_i$,  where $f^\dag_{i\sigma}$ is the spinon creation operator and $h_i$ is the holon annihilation operator.  This decomposition {is valid} when the local constraint  {is enforced for every site $i$ by}
\begin{eqnarray}
	G_i=h_i^\dag h_i+\sum_\sigma f_{i\sigma}^\dag f_{i\sigma}-1=0.
\end{eqnarray}
The mean field Hamiltonian of the $t$-$J$ model in the slave boson {representation} reads~\cite{NLPRL,LN}, 
	\begin{eqnarray}
		L_{MF}&=&\frac{J}4 \sum_{\langle ij\rangle}[|\gamma^f|^2+|\Delta_a|^2-\sum_\sigma (\gamma^{f\dag}{\rm e}^{{\rm i} a_{ij}}f^\dag_{i\sigma} f_{j\sigma}+h.c.)]\nonumber\\
		&+&\sum_{\langle ij\rangle}\frac{J}4[\Delta_a{\rm e}^{{\rm i}\phi_{ij}}(f_{i\uparrow}^\dag f_{j\downarrow}^\dag-f_{i\downarrow}^\dag f_{j\uparrow}^\dag)+h.c.]\nonumber\\
		&+&\sum_ih^\dag_i(\partial_\tau-\mu_h) h_i+\sum_{i\sigma}f^\dag_{i\sigma}(\partial_\tau -\mu_f)f_{i\sigma}\nonumber\\
		&-&t\sum_{\langle ij\rangle}({\rm e}^{{\rm i}a_{ij}}(\gamma^f h^\dag_i h_j+\gamma^{h\dag} f^\dag_{i\sigma} f_{j\sigma})+h.c.)\nonumber\\
		&+&\sum_i {\rm i}g\lambda_iG_i, \label{LMF}
	\end{eqnarray}
	where $\Delta_a$ for $a=x,y$ labels the pairing parameter in the $a$-link, and $\gamma ^{h,f}$ are the hopping parameters for the spinon and the holon.   {We} choose $\Delta_a$ and $\gamma ^{h,f}$  {as} expectation values in the mean field approximation. The phase fields $a_{ij}$ and $\phi_{ij}$, which obey the periodic boundary condition, are  quantum fluctuations {introduced} to compensate  {for} the gauge symmetry breaking due to the mean field approximation.  Besides the temporal gauge field  $\lambda_i$,  the mean field theory has a spatial gauge invariance under the transformations $(f_{i\sigma}(\tau),h_i(\tau))\to {\rm e}^{i\theta_i(\tau)}(f_{i\sigma}(\tau),h_i(\tau))$, and $a_{ij}\to a_{ij}+\theta_i-\theta_j$ and $\phi_{ij}\to  \phi_{ij}+\theta_i+\theta_j$. 
  The equation of motion of $a_{ij}$ leads to the constraint of vanishing counterflow 
  between the holon and spinon currents {due to the mean field approximations}~\cite{IL}, i.e., 
	\begin{eqnarray}
		J_{ij}=J^f_{ij}+J^h_{ij}=0.\label{cconstraint}
	\end{eqnarray} 
	This is also a local constraint.  
	 We have shown that the variation of $\phi_{ij}$  does not result in new constraints \cite{LYL}.
	Therefore, our problem {is} how to quantize the mean field theory with the constraints $G_i=0$ and $J_{ij}=0$ 
	{under} proper gauge fixing conditions. \\

\section{Brief Review on our previous work \cite{LYL}}

We briefly review the main results obtained in \cite{LYL}. In the continuum limit, the mean field Lagrangian \eqref{LMF} with a $d$-wave pairing is given by
\begin{eqnarray}
		L_{MFC}&=&\int {\rm{d}}^2r [ \sum_{\sigma}f^\dag_{\sigma}(\partial_\tau-\mu_f-{\rm i}g \delta \lambda)f_\sigma\nonumber\\
		&+&h^\dag(\partial_\tau-\mu_h-{\rm i}g \delta \lambda)h]\nonumber\\
		&-&\int {\rm{d}}^2r[\frac{1}{2m_f} \sum_{\sigma,a}f^\dag_{\sigma}(-{\rm i}\partial_a-g \delta a_a)^2f_\sigma\nonumber\\
		&-&\frac{1}{2m_h}\sum_ah^\dag(-{\rm i}\partial_a-g \delta a_a)^2h]\nonumber\\
		&+&\frac{1}2\int {\rm d}^2r \sum_a (\Delta_a \partial_a({\rm e}^{{\rm i}\phi/2}f^\dag_\uparrow) \partial_a({\rm e}^{{\rm i}\phi/2}f^\dag_\downarrow)+h.c.), \label{LMFC}\nonumber\\
	\end{eqnarray}
where {the uniform RVB state is taken in} the mean field state of the gauge field  {due to time-reversal symmetry} \cite{Anderson}.  The pure gauge field  Lagrangian is $L_0(a_\mu)=0$ where $a_\mu=(a_\tau, a_i)=(\delta\lambda,\delta a_i)$. To retain the constraints and remove the gauge redundancy, we  tried the BRST invariant gauge fixing term as  
\begin{eqnarray}
		L_{GFL}&=&-\int {\rm d}^2r {\frac{1}{2\xi}} (\zeta\partial_\tau\delta\lambda+\sum_a\partial_a \delta a_a)^2\nonumber\\
		&+&\int {\rm d}^2r\bar u(\zeta\partial_\tau^2+\sum_a\partial^2_a)u, \label{eq6}
	\end{eqnarray}
	where {$\xi$ is an arbitrary (gauge) parameter and $\zeta^{-1}$ is similar to the speed of light}. This is called the Lorenz gauge fixing and is invariant under the BRST transformation
	\begin{eqnarray}
		&&\delta_{B}f_\sigma={\rm i}\epsilon gu f_\sigma, \delta_{B}h={\rm i}\epsilon gu h,\delta_B\phi=2{\rm i}\epsilon g u,\nonumber\\
		&&\delta_{B} \delta\lambda=\epsilon \partial_\tau u,\delta_B \delta a_b=\epsilon \partial_bu, \delta_{B}u=0,\nonumber\\
		&& \delta_B\bar u=\epsilon\xi^{-1} (\zeta\partial_\tau\delta\lambda+\sum_b\partial_b\delta a_b).  \label{BRSTT}
	\end{eqnarray}
The BRST transformation is nilpotent, i.e, $\delta^2_B=0$. ($\delta^2_B\bar u=0$ due to the equation of motion for $u$. We here use the on-shell BRST transformation.  {Notice that Eq.~\eqref{eq6} can be obtained by a so-call off-shell BRST exact form (see e.g.,~\cite{Weinberg}). Defining $\delta_B=\epsilon s$, we introduce the off-shell BRST Lagrangian 
\begin{eqnarray}
    L^{off}_{GFL}=s[\bar u( {\frac{\xi}2}\Pi-(\zeta\partial_\tau\delta\lambda+\sum_b\partial_b\delta a_b))],
\end{eqnarray}
and the BRST transformations are modified as 
\begin{eqnarray}
{\delta}_B \bar u=\Pi,\quad \delta_B\Pi=0,
\end{eqnarray}
while other fields' are the same as those in Eq.~\eqref{BRSTT} . The auxiliary field $\Pi$ is the Nakanishi–Lautrup field.  This off-shell gauge fixing term is explicitly BRST exact. The Euler-Lagrange equation of $\Pi $ given  {$\Pi=\xi^{-1}(\zeta\partial_\tau\delta\lambda+\sum_b\partial_b\delta a_b)$} and substituting  which into $L^{off}_{GFL}$, one recovers Eq.~\eqref{eq6}.    
}   However, it is easy to check that the free {propagator} of the gauge field is ill-defined 
because $ D^{(0)-1}_{\mu\nu}(i\omega,{\boldsymbol q})\propto q_\mu q_\nu$,  so that $\det D^{(0)-1}=0$.

To overcome this difficulty and get a well-defined free gauge propagator, we introduced another  gauge fixing term 
\begin{eqnarray}\label{Eq:LGFgh}
		L_{GF}&=&\frac{A}{2}(\partial_\tau\delta\lambda)^2+B\sum_b \partial_\tau \delta a_b\partial_b\delta \lambda
		+\frac{C}{2}(\sum_b \partial_b \delta a_b)^2\nonumber\\
		&+&\frac{D}{2}\sum_b (\partial_\tau \delta a_b)^2+\frac{E}{2}\sum_b(\partial_b \delta\lambda)^2+\bar{u}K u, \label{eq8}
\end{eqnarray}
where
\begin{eqnarray}
			&& {A=C\zeta^2,\quad C\zeta=B+D=B+E,\quad E=D},
        \label{Eq:ABCDE_relation}
\\
			&&K=- {C\xi}(\zeta\partial^2_\tau +\sum_b\partial_b^2).
	\end{eqnarray} 
This gauge fixing term is also the BRST invariant.  {Furthermore, when $D=E=0$, Eq. (\ref{Eq:ABCDE_relation}) implies $A/\zeta^2=B/\zeta=C$, which reduces to the Lorenz gauge for $C=1/\xi$.}  It is easy to {verify} that the free propagator of the gauge field is well-defined except {when} $D=E=0$.  {In our previous work \cite{LYL}, we set $B=0$ while keeping $D,E\neq 0$ to decouple  the temporal and spatial components  for convenience}.  The perturbation calculations in~\cite{LYL} are all based on $L_{MFC}+L_{GF}$. 	


\section{Flaws in the  {previous} section and a new point of view \label{Qu}} 
 
\subsection{The BRST exact form of the gauge fixing term}

We added a BRST invariant term Eq. \eqref{eq8} to $L_0(a_\mu)=0$ as the gauge fixing. However, we did not prove {whether} the added gauge fixing 
term is {of} the BRST exact form or not. According to the BRST cohomology, only when the added gauge fixing term is
 BRST exact  {does it} not introduce {extra} dynamics to the system. For example, the Maxwell term is BRST invariant but not BRST exact. If we add this term with a Lorenz gauge fixing to $L_0(a_\mu)=0$, extra gauge field dynamics is introduced to the system.
 
 Let us {determine} what kind of gauge fixing terms are BRST exact.  We consider a general BRST {transformation} 
\begin{eqnarray}
    \delta_B a_\mu={\epsilon\partial_\mu u}, \delta_B u=0, {\delta_B} \bar u= \epsilon F(a_\mu),
\end{eqnarray}
with $F(a_\mu)$ to be determined. 
Denoting $\delta_B=\epsilon s$, a BRST exact form which is added to the {Lagrangian} is assumed to be proportional to  
\begin{eqnarray}
    s(\bar u \sum_{\mu \nu}C_{\mu\nu}\partial_\mu a_\nu)=F\sum_{\mu \nu}C_{\mu\nu}\partial_\mu a_\nu
    -\bar u \sum_{\mu \nu}C_{\mu\nu}\partial_\mu \partial_\nu u.~~
\end{eqnarray}
The equation of motion for $u$ is $\sum_{\mu \nu}C_{\mu\nu}\partial_\mu \partial_\nu u=0$. This leads to
\begin{align}
F(a_\mu)={\frac{1}{2\xi}}\sum_{\mu,\nu}C_{\mu\nu}\partial_\mu a_\nu,\label{Eq:F}
\end{align}
 due to the requirement of $s^2=0$. 
 Therefore,  {a general BRST exact term  is given by 
\begin{eqnarray}
	L_{GGF}&=&{\frac{1}{2\xi}}(\sum_{\mu \nu}C_{\mu\nu}\partial_\mu a_\nu)^2
    -\bar u \sum_{\mu \nu}C_{\mu\nu}\partial_\mu \partial_\nu u.~~\label{GGF}
\end{eqnarray}
The Lorenz gauge is recovered by taking 
\begin{align}
    C_{\mu\nu}=\delta_{\mu\nu}.
\end{align}
As before, Eq.~\eqref{GGF} can be obtained from a BRST exact off-shell  Lagrangian
\begin{eqnarray}
L^{off}_{GGF}&=&s[\bar u ( \frac\xi{2}\Pi-\sum_{\mu \nu}C_{\mu\nu}\partial_\mu a_\nu)],
\end{eqnarray}}

The inverse of the free gauge propagator corresponding to Eq.~\eqref{GGF} can be read out, $D^{(0)-1}\propto\tilde k_\mu\tilde k_\nu$ for $\tilde k_\mu=\sum_\rho C_{\mu\rho}k_\rho$. 
Therefore the free propagator of the gauge field $D^{(0)}$ now is ill-defined  
because $\det(D^{(0)-1})\propto \det(\tilde k_\mu\tilde k_\nu)=0$.

 On the other hand,  {the off-shell version of {Eq.~\eqref{eq8}} can be written as {(up to a surface term)}  
 {
\begin{eqnarray}\label{Eq:LGFgh}
		L^{off}_{GF}&=&\frac {1}{2}(A-\frac{\zeta^2}{\xi})(\partial_\tau\delta \lambda)^2+(B-\frac{\zeta}{\xi})(\sum_b\partial_b\delta a_b)
        (\partial_\tau\delta\lambda)
		\nonumber\\
		&+&\frac{1}{2}(C-\frac{1}{\xi})\sum_b(\partial_b\delta a_b)^2\nonumber\\
        &+&\frac{D}{2}\sum_b (\partial_\tau \delta a_b)^2+\frac{E}{2}\sum_b(\partial_b \delta\lambda)^2\nonumber\\
		&+&Cs[\bar u ( {\frac{\xi}{2}}\Pi-(\zeta\partial_\tau\delta\lambda+\sum_b\partial_b\delta a_b))].\label{LoffGF}
\end{eqnarray}}
 The last term is a BRST exact form. The rest terms are actually gauge invariant and thus are not of the BRST exact form. Therefore, $L^{off}_{GF}$ is not BRST exact.  Obviously, when $E=D=0$  {and $C=1/\xi$}, $L^{off}_{GF}$ recovers $L^{off}_{GFL}$. Thus, the gauge invariant terms in $L^{off}_{GF}$ can be thought as the regulator to ensure a well-defined full propagator of the gauge field.}  
\\

 \subsection{The exact local constraint problem}
 
 The BRST symmetry is a global symmetry; Noether's theorem gives the conserved BRST charge $Q$ which counts 
 the ghost {number} of a state.  {If  we denote the BRST exact  term in Eq. \eqref{LoffGF} as $s\Psi$ which serves as the gauge fixing term and 
 a small {variation} of such a gauge fixing condition as $s\tilde \delta\Psi$, the {variation} of any {physical} matrix element vanishes \cite{Weinberg}, i.e.,  
	\begin{eqnarray}
	\tilde\delta \langle \alpha|\beta\rangle=-\langle\alpha|\tilde\delta \int {\rm d}\tau L|\beta\rangle=-\langle\alpha|\{Q^{off},\tilde\delta\Psi\}|\beta\rangle=0.
	\end{eqnarray}
	Since $\tilde \delta \Psi$ is arbitrary, one must have
	\begin{eqnarray}
	\langle\alpha| Q^{off}=0,~ Q^{off}|\beta\rangle=0.
	\end{eqnarray}
    Integrating over the auxiliary field, this proves the BRST charge free requirement of the physical states, $$Q|phys\rangle=0.$$ }
 For the theory with the gauge fixing term  in Eq. \eqref{Eq:LGFgh}, the on-shell BRST charge is given by \cite{LYL} 
 \begin{eqnarray}
		Q&=&\int {\rm{d}}^2x~ ({\rm i}gG +E\sum_b\partial_b^2\delta\lambda-D\sum_b\partial_\tau\partial_b \delta a_b )u\nonumber\\
		&+&[A\partial_\tau\delta\lambda +(B+D)\sum_b\partial_b \delta a_b]\partial_\tau u.\label{eq21}
	\end{eqnarray}
The constraints read out from the BRST charge free of the physical states are given by 
 \begin{eqnarray}
		&& ({\rm i}gG +E\sum_b\partial_b^2\delta\lambda-D\sum_b\partial_\tau\partial_b \delta a_b )=0,\nonumber\\
		&&[A\partial_\tau\delta\lambda +(B+D)\sum_b\partial_b \delta a_b]=0.
	\end{eqnarray}
	While the second equation is the gauge fixing condition, the first one does not exactly give the local constraint $G=0$ {unless} both $E$ and $D$ are zero. 
	In deriving the conserved charge from Noether's theorem, the Euler-Lagrange equations of the fields are used. {For example,} the equations of motion of the gauge field $\delta a_i$ gives 
	\begin{eqnarray}
		J_b(\delta  a) =B\partial_\tau\partial_b\delta\lambda+D\partial^2_\tau \delta a_b+C\partial_b(\sum_c\partial_c \delta a_c), \label{Jne0}
	\end{eqnarray} 
	which also does not  {yield} the constraint $J_b=0$.

\subsection{Completeness of consistent gauge theory}

In {light of the} discussions in Sec. \ref{Qu},  our theory in \cite{LYL} {still has} some problems: 
{(1)} {Extra} artificial gauge field dynamics was introduced. (2) Both {the} local number and current constraints are not
exactly {recovered}. The above two problems can be {solved by applying} the Lorenz gauge {in the $D=E=0$ limit}. On the one hand, the gauge fixing term 
is BRST exact so that no extra dynamics is introduced by hand; on the other hand, the BRST charge becomes 
\begin{eqnarray}
		Q=\int \mathrm{d}^2x~ ({\rm i}gG u+{B}[{\zeta}\partial_\tau\delta\lambda +\sum_b\partial_b \delta a_b]\partial_\tau u). \label{Qcon}
	\end{eqnarray}
	This gives the local constraint: $G=0$ and the Lorenz gauge ${\zeta}\partial_\tau\delta\lambda +\sum_b\partial_b 
	\delta a_b=0$. The vanishing counterflow  constraint becomes
	 \begin{eqnarray}
		J_b(\delta  a) ={C\partial_b}({\zeta}\partial_\tau\delta\lambda+\sum_c\partial_c \delta a_c)=0, \label{Jcon}
	\end{eqnarray}   
	where the second equality comes from the Lorenz gauge because {of Eq. \eqref{Eq:ABCDE_relation}}. However, as we have shown that the free propagator 
	of the {gauge} field is ill-defined for the Lorenz gauge. The perturbation calculation {therefore} seems to be {invalid}. {Instead of Eq.~\eqref{eq8}, one may try to add a Maxwell term to the gauge field as  the usual treatment in quantum electrodynamics where the propagator of the gauge field is well-defined. But, there will be extra contributions to the BRST charge \eqref{eq21} and the current \eqref{Jne0} stemming from the Maxwell term (because the Maxwell term is not BRST-exact) unless the strength of the Maxwell term goes to zero, and the problem of ill-defined propagator remains.} 

	Can we have a solution to these problems? Let {us} recall the source of the {ill-defined nature} of the free propagator of the gauge field.
	As we have observed, the zero mode in the  gauge field actually comes from the  {skew} $U(1)$ gauge symmetry 
	of the Lorenz gauge fixing term \cite{LYL}, i.e., when $$a_\mu\to a_\mu+\epsilon_{\mu\nu}\partial_\nu\theta,$$ 
    the Lorenz gauge fixing condition is invariant, 
	$$\partial_\mu a_\mu\to \partial_\mu( a_\mu+\epsilon_{\mu\nu}\partial_\nu\theta)= \partial_\mu a_\mu.$$ 
	This leads to {the situation where} the inverse of the free propagator of the gauge field, $D^{(0)}_{\mu\nu}\propto q_\mu q_\nu$ has a vanishing determinant. 
However, this skew $U(1)$ gauge invariance holds only for  the gauge fixing term.
The whole Lagrangian $L_{MFC}+L_{GFL}$ is not  {skew} gauge invariant because the minimal coupling between the gauge field and the matter field 
breaks this symmetry. 
Thus, the full propagator $D_{\mu\nu}$ is well-defined.  To avoid the divergence of $D^{(0)}_{\mu\nu}$, 
we use the Lagrangian $L_{GF}$, i.e., \eqref{Eq:LGFgh},  in order to keep the BRST symmetry of the theory, and take $D=E\to 0$ limit after {the} calculations. 
In this case, if we denote the gauge field propagator as $D'_{\mu\nu}$, then the free propagator $D^{'(0)}_{\mu\nu}$ is invertible. According to Dyson's equation, the full propagator is
\begin{eqnarray}
D'_{\mu\nu}=[(D'^{(0)-1}-\Pi'^*)^{-1}]_{\mu\nu},
\end{eqnarray}
where $\Pi'^*$ is the vacuum {polarization} of the holon $h$ and the spinon $f_\sigma$. When $D,E\to 0$, we have the well-defined 
\begin{eqnarray}
D_{\mu\nu}=[(D^{(0)-1}-\Pi^*)^{-1}]_{\mu\nu}.
\end{eqnarray} 
For example, if we take the random phase approximation (RPA) for the propagator, one has
\begin{eqnarray}
D'^{RPA}_{\mu\nu}=[(D'^{(0)-1}-\Pi'^{(0)*})^{-1}]_{\mu\nu}.
\end{eqnarray}
where $\Pi'^{(0)*}=\Pi^{(0)*}$ does not have the contribution from the gauge field propagator. And
 $D'^{RPA}\to D^{RPA} $ when $D,E\to0$  {(see Appendix \ref{app2})}. We then have a  well-defined 
$D^{RPA}$.  {The same process also works for adding a Maxwell term and then taking the zero strength limit.} 

We can {replace} $D^{(0)}$ {with} $D^{RPA}$ in all calculations. In this way, we are able to {perform} 
a perturbative calculation {for} the {consistent} gauge theory in the slave boson {representation} of the $t$-$J$ model 
in the strong coupling {limit}.  

\section{conclusions}

We now {complete the construction of} the consistent gauge theory {for the $U(1)$ slave particle representation of the $t$-$J$ model. We demonstrated that the gauge fixing condition is BRST-exact and show that  local constraints are exactly preserved in the Lorenz gauge. The difficulty of ill-defined free gauge propagator was also resolved. This work thus establishes a concrete framework for systematic perturbation theory.} The further tasks will be to perform the hard work of
the perturbation calculations for the physical observables.  {For example, the first task is to re-calculate the key observables in the strange metal phase from \cite{LYL} using the new, consistent formalism presented here.} {We provide a consistent formalism that duals a strongly correlated system into a controllable weakly coupled gauge theory with constraints, this may pave a way towards a deeper understanding of the physics in other strongly correlated systems, such as the overdoped regime of the superconducting phase and the pseudo gap regime of cuprates, Mott physics, and spin liquids. }

\acknowledgements

The authors thank  Q. Niu for  useful discussions. This work is supported by the National Natural
Science Foundation of China with Grants No.~12174067 (X.L.
and Y.Y.), No.~12204329 (L.L.),  {No.~12135018 (T.S.), and No.~12047503 (T.S.). T.S. is also supported by National Key Research and Development Program of China with Grant No.~2021YFA0718304, and by CAS Project for
 Young Scientists in Basic Research with Grant No.~YSBR-057}.


\begin{widetext} 

\appendix

\section{BRST symmetry on lattice} \label{app1}

For the completeness of this paper, we write some results of the lattice version of consistent gauge theory with the BRST symmetry in this appendix.  

For the t-J model, the partition function on lattice reads
\begin{equation}
    Z=\int D f D f^\dag D b D b^* D \lambda D\chi D\Delta \exp \left( -\int_0^\beta {\rm d}\tau L_1 \right),
\end{equation}
with 
\begin{eqnarray}
      L_1&=&\tilde{J} \sum_{\langle ij\rangle} (|\chi_{ij}|^2+|\Delta_{ij}|^2)+\sum_{i,\sigma}f^{\dag}_{i\sigma} (\partial_\tau-\mu_f)f_{i\sigma}+\sum_i b^*_i(\partial_\tau-\mu_b)b_i-\sum_{ij} {t_{ij}}b_ib_j^*f_{i\sigma}^\dag f_{j\sigma}\nonumber\\
      &&-{\rm i}\sum_i \lambda_i(f_{i\sigma}^\dag f_{i\sigma}+b^*_i b_i-1)\nonumber\\
      &&-\tilde{J}\left[ \sum_{\langle ij\rangle} \chi^*_{ij} \left( \sum_\sigma f^\dag_{i\sigma} f_{j\sigma}+c.c.\right)\right]+\tilde{J}\left[\sum_{\langle ij\rangle} \Delta_{ij}(f^\dag_{i\uparrow}f^\dag_{j\downarrow}-f^\dag_{i\downarrow}f^\dag_{j\uparrow})+c.c. \right],
\end{eqnarray}
  with 
  \begin{equation}
      \tilde{J}=J/4, \quad \chi_{ij}=\sum_\sigma( f^\dag_{i\sigma}f_{j\sigma} ), \quad \Delta_{ij}=( f_{i\uparrow}f_{j\downarrow}-f_{i\downarrow}f_{j\uparrow} ).
  \end{equation}
There is a gauge invariance,
\begin{equation}
    f_{i\sigma}\rightarrow {\rm e}^{{\rm i}\theta_i}f_{i\sigma}, \quad 
    b_{i}\rightarrow {\rm e}^{{\rm i}\theta_i}b_{i}, \quad
    \chi_{ij} \rightarrow {\rm e}^{-{\rm i}\theta_i}\chi_{ij} {\rm e}^{{\rm i}\theta_j}, \quad
    \Delta_{ij} \rightarrow {\rm e}^{{\rm i}\theta_i}\Delta_{ij} {\rm e}^{{\rm i}\theta_j}, \quad
    \lambda_i\rightarrow \lambda_i+\partial_\tau \theta_i.
\end{equation}
In the mean field theory, one chooses $\chi_{ij}=\sum_\sigma\langle f^\dag_{i\sigma}f_{j\sigma} \rangle$ and $\Delta_{ij}=\langle f_{i\uparrow}f_{j\downarrow}-f_{i\downarrow}f_{j\uparrow} \rangle$. In the uniform RVB mean field theory, one assume
\begin{equation}
    \chi_{ij}=\chi=real.
\end{equation}
Here we neglect the $\Delta$ field and consider the $\chi$
 and
$\lambda$ fields. There are amplitude and phase fluctuations of
the $\chi$
 field, but the amplitude fluctuations are massive
and do not play an important role in the low-energy
limit.  {Furthermore, the mean field ground state we consider is the uniform RVB state, which is topologically trivial. Then, the ghost zero modes are ignored.} Therefore the relevant Lagrangian to start with is
\begin{equation}
    L_0=\sum_{i\sigma}f^\dag _{i\sigma}(\partial_\tau -\mu_f+{\rm i}a_0(r_i))f_{i\sigma}+
    \sum_{i}b^* _{i }(\partial_\tau -\mu_b+{\rm i}a_0(r_i))b_{i }-\tilde{J}\chi 
    \sum_{\langle ij\rangle,\sigma} ({\rm e}^{{\rm i}a_{ij}}f^\dag _{i\sigma}f_{j\sigma}+h.c.)-t\chi \sum_{\langle ij\rangle}({\rm e}^{{\rm i}a_{ij}}b^* _{i }b_{j }+h.c.),
\end{equation}
with
\begin{equation}
    a_{ij}\rightarrow a_{ij}+\theta_i-\theta_j, \quad a_0(i)\rightarrow a_0(i) +\partial_\tau \theta_i(\tau).
\end{equation}
 {The equation of motion of $a_{ij}$ leads to the constraint of vanishing counterflow between the holon and spinon currents}, i.e.,
	\begin{eqnarray}		J_{ij}=J^f_{ij}+J^h_{ij}=0.\label{cconstraint}
	\end{eqnarray} 
	This is also a local constraint. {Note that the gauge field appears  in the expression of the spinon and holon currents. And the constraint holds for non-vanishing gauge configurations.} The variation of $\phi_{ij}$  does not result in new constraints (Higgs mechanism).

\subsection{Gauge fixing term Eq. \eqref{eq8} on lattice}

  {We abbreviate $\delta \lambda$ and $\delta a_a$ as $\lambda$ and $a_a$ for convenience.} The gauge fixing terms \eqref{eq8} are
\begin{eqnarray}
L_{GF}&=&\frac{A}{2}(\partial_\tau a_\tau)^2+\sum_a \frac{D}{2}(\partial_\tau a_a)^2+\sum_a \frac{E}{2}(\partial_aa_\tau)^2+\sum_{ab}\frac{C}{2}(\partial_aa_a)(\partial_ba_b)+\sum_a{B}(\partial_\tau a_a)
(\partial_aa_\tau)+ {L_{gh}},\\
L_{gh}&=&\bar{u}Ku,
\end{eqnarray}
where $\mu=\tau,a$, $a,b=x,y$ for 2 spatial dimensions.

 The lattice version for $a_\tau({\boldsymbol r},\tau)$ is just $\lambda_i(\tau)$ for $i$, the two-dimensional lattice index. $a_a({\boldsymbol r})$ comes from the $U_{ij}={\rm e}^{iga_{ij}}$. The spatial derivative is given by $\frac{\partial f}{\partial x_{\hat \delta}}\to \Delta_{i,\hat\delta}f=f_{i+\hat\delta}-f_i$. $\Delta^+_{i,\hat\delta}$ means $\hat\delta=\hat x,\hat y$ in the forward direction. Thus,  
\begin{eqnarray}
\partial_aa_0\to \lambda_{i+\hat\delta}-\lambda_i.
\end{eqnarray}
On the other hand,  $a_{ij}=({\boldsymbol r}_j-{\boldsymbol r}_i)\cdot {\boldsymbol a}(\frac{{\boldsymbol r}_i+{\boldsymbol r}_j}2)$ .

In the usual treatment of the BRST fermionic ghost, $u_i$ lives on lattice sites. 
\begin{equation}
    \int {\rm d}^2r\bar u(\zeta\partial_\tau^2+\sum_a\partial^2_a)u\rightarrow  \sum_{i}  \bar{u}_i(\partial_\tau^2+\Delta^2)u_i,
\end{equation}
where $\Delta^2$ is the Laplacian on lattice in  {two-dimensions},
$
  \Delta^2 u_i=\sum_{{\boldsymbol e}_j}(u_{i+{\boldsymbol e}_j}+u_{i-{\boldsymbol e}_j}-2u_i)  
$,  {and ${\boldsymbol e}_j$ is the unit vector.}

\subsection{Checking lattice BRST symmetry}
 The Fourier transformation is defined as $f_i=\sum_n\int {\rm d}^2k f_{\omega_n,k}{\rm e}^{\omega_n\tau+{\rm i}\Vec{k}\cdot \Vec{r}}$, and  $a_{i,i+{\boldsymbol e}_j}=\int {\rm d}^2 ka_k{\rm e}^{{\rm i}k(i+i+{\boldsymbol e}_j)/2}$, where we assume all the gauge fields are real. The Lagrangian we are now considering is
\begin{eqnarray}
    L&=&L_0+L_{gf}+L_{gh},\\
    L_0&=&
    \sum_{i\sigma}f^\dag _{i\sigma}(\partial_\tau -\mu_f+{\rm i}\lambda_i)f_{i\sigma}+
    \sum_{i }b^* _{i }(\partial_\tau -\mu_b+{\rm i}\lambda_i)b_{i }-\tilde{J}\chi 
    \sum_{\langle ij\rangle,\sigma} ({\rm e}^{{\rm i}a_{ij}}f^\dag _{i\sigma}f_{j\sigma}+h.c.)-t\chi \sum_{\langle ij\rangle}({\rm e}^{{\rm i}a_{ij}}b^* _{i }b_{j }+h.c.)\nonumber\\
    &=&
    \sum_{i\sigma}f^\dag _{i\sigma}(\partial_\tau -\mu_f+{\rm i}\lambda_i)f_{i\sigma}+
    \sum_{i }b^* _{i }(\partial_\tau -\mu_b+{\rm i}\lambda_i)b_{i }\nonumber\\
    &&-\tilde{J}\chi 
    \sum_{\langle ij\rangle,\sigma} ((1+{\rm i}a_{ij}-\frac{a_{ij}^2}{2}+\mathcal{O}(a^3))f^\dag _{i\sigma}f_{j\sigma}+h.c.)-t\chi \sum_{\langle ij\rangle}((1+{\rm i}a_{ij}-\frac{a_{ij}^2}{2}+\mathcal{O}(a^3))b^* _{i }b_{j }+h.c.)\nonumber\\
    &=&  \int {\rm d}^2 {\boldsymbol k}d^2{{\boldsymbol p}}d^2{{\boldsymbol q}} (\sum_{\sigma}((\omega_n-\mu_f)f^\dag_{k\sigma}f_{k\sigma}+{\rm i}\lambda_pf^\dag_{k+p,\sigma}f_{k\sigma})+((\nu_n-\mu_b)b^*_{k}b_{k}+{\rm i}\lambda_pb^\dag_{k+p}b_k)\nonumber\\
    &&-\tilde{J}_\chi (2\sum_{{\boldsymbol e}_j}\cos({\boldsymbol {k\cdot e}}_j) f_{k\sigma}^\dag f_{k\sigma}+2\sum_{{\boldsymbol e}_j}\sin({( {\frac{\boldsymbol k}{2}+\boldsymbol{p}})\cdot} {\boldsymbol e}_j)a_kf^{\dag}_{k+p,\sigma}f_{p\sigma}-\sum_{{\boldsymbol e}_j}\cos({ (\frac{\boldsymbol k}{2}+\frac{\boldsymbol p}{2}+\boldsymbol{q})}\cdot {\boldsymbol e}_j)a_k a_p f^\dag_{k+p+q,\sigma}f_{q\sigma})\nonumber\\
    &&-t_\chi(2\sum_{{\boldsymbol e}_j}\cos({\boldsymbol {k\cdot e}}_j) b_{k}^* b_{k}+2\sum_{{\boldsymbol e}_j}\sin({ (\frac{\boldsymbol k}{2}+\boldsymbol p)\cdot} {\boldsymbol e}_j)a_kb^{*}_{k+p}b_{p}-\sum_{{\boldsymbol e}_j}\cos({ (\frac{\boldsymbol k}{2}+\frac{\boldsymbol p}{2}+\boldsymbol q)}\cdot {\boldsymbol e}_j)a_k a_p b^*_{k+p+q}b_{q}) ,\\
    L_{gf}
    &=&\sum_i(  \frac{A}2
    (\partial_\tau\lambda_i)^2+\frac{D}{2}\sum_{{\boldsymbol e}_j}(\partial_\tau a_{i,i+{\boldsymbol e}_j})^2+\frac{E}{2}\sum_{{\boldsymbol e}_j}(\lambda_{i+{\boldsymbol e}_j}-\lambda_i)^2+\frac{C}{2 } \sum_{{\boldsymbol e}_j}( a_{i,i+{\boldsymbol e}_j}-a_{i-{\boldsymbol e}_j,i})^2\nonumber\\
    &+&B\sum_{{\boldsymbol e}_j} (\partial_\tau a_{i,i+{\boldsymbol e}_j}(\lambda_{i+{\boldsymbol e}_j}-\lambda_i)))\nonumber\\
    &=&\sum_i  (\frac{A}2
    (\partial_\tau\lambda_i)^2+\frac{D}{2}\sum_{{\boldsymbol e}_j}(\partial_\tau a_{i,i+{\boldsymbol e}_j})^2+\frac{E}{2}\sum_{{\boldsymbol e}_j}(\Delta_{i,{\boldsymbol e}_j}^+\lambda_i)^2+\frac{C}{2 } \sum_{{\boldsymbol e}_j}( \Delta_{i,{\boldsymbol e}_j}^-a_{i,i+{\boldsymbol e}_j})^2 +B\sum_{{\boldsymbol e}_j} (\partial_\tau a_{i,i+{\boldsymbol e}_j}\Delta_{i,{\boldsymbol e}_j}^+\lambda_i))\nonumber\\
    &=&\int {\rm d}^2 k(\frac{A \nu_n^2 \lambda_k\lambda_{-k}}{2}+\frac{D}{2}(\nu_n^2 a_{k }a_{-k} )+\frac{E}{2}\sum_{{\boldsymbol e}_j}(2-2\cos({\boldsymbol k\cdot \boldsymbol e}_j))\lambda_k\lambda_{-k}+\frac{B}{2}(\nu_n (\exp({\rm i}({\boldsymbol k\cdot \boldsymbol e}_j))-1) a_k \lambda_{-k}+h.c.)\nonumber\\
        &&+\frac{C}{2}(a_{-k_x},a_{-k_y})
        \left[
		\begin{array}{cccc}
			2-2\cos({\boldsymbol k\cdot \boldsymbol e}_x) &4\sin(\frac{{\boldsymbol k\cdot \boldsymbol e}_x}{2})\sin(\frac{{\boldsymbol k\cdot \boldsymbol e}_y}{2})  \\
			4\sin(\frac{{\boldsymbol k\cdot  \boldsymbol e}_x}{2})\sin(\frac{{\boldsymbol k\cdot \boldsymbol e}_y}{2}) & 2-2\cos({\boldsymbol k\cdot \boldsymbol e}_y)  
		\end{array}
		\right]
        \left(
        \begin{array}{cccc}
			a_{k_x}\\
			a_{k_y}
		\end{array}
        \right)
        )\\
    L_{gh}&=&  {-C} \sum_{i}  \bar{u}_i( {\zeta}\partial_\tau^2+\Delta^2)u_i= {C} \int {\rm d}^2k ( {\zeta} \omega_n^2 +2\sum_{{\boldsymbol e}_j}(\cos({\boldsymbol k\cdot \boldsymbol e}_j)-1))\bar{u}_k u_k,
\end{eqnarray}
where $\lambda_{-k}=\lambda^\dag_k$, and $a_{-k}=a_{k}^\dag$.

The BRST transformations are:
\begin{eqnarray}
 &&\delta_{B}f_{i,\sigma}=-{\rm i}\epsilon gu_i f_{i,\sigma}, \delta_{B}h_i=-{\rm i}\epsilon gu_i h_i, \delta_{B}  \lambda_i=\epsilon \partial_\tau u_i,\nonumber\\
 &&\delta_B   a_{i,i+{\boldsymbol e}_j}=\epsilon \Delta^+_{i,{\boldsymbol e}_j}u_i=\epsilon(u_{i+{\boldsymbol e}_j}-u_i), \delta_{B}u_i=0,\nonumber\\
&& \delta_B\bar u_i= {\frac{1}{\xi}}(\zeta\partial_\tau \lambda_i+\sum_{{\boldsymbol e}_j}  \Delta_{i,{\boldsymbol e}_j}^-a_{i,i+{\boldsymbol e}_j})= {\frac{1}{\xi}}(\zeta\partial_\tau \lambda_i+\sum_{{\boldsymbol e}_j}  (a_{i,i+{\boldsymbol e}_j}-a_{i-{\boldsymbol e}_j,i})).
\end{eqnarray}
Let's check
\begin{eqnarray}
\delta_BL_{gh}&=& {-C\xi}\sum_i \delta_B\bar{u}_i( {\zeta}\partial_\tau^2+\Delta^2)u_i\nonumber\\
&=& {-\epsilon C}\sum_i  ( {\zeta}\partial_\tau \lambda_i+\sum_{{\boldsymbol e}_j}  \Delta_{i,{\boldsymbol e}_j}^-a_{i,i+{\boldsymbol e}_j})( {\zeta}\partial_\tau^2+\Delta^2)u_i,\nonumber\\
&=& {-\epsilon C}\sum_i (
{ {\zeta^2} \partial_\tau \lambda_i \partial_\tau ^2 u_i}
+
{ {\zeta}\partial_\tau \lambda_i \Delta^2 u_i}\nonumber\\
&&+
 {\zeta}{(\sum_{{\boldsymbol e}_j} \Delta_{i,{\boldsymbol e}_j}^-a_{i,i+{\boldsymbol e}_j})\partial_\tau^2 u_i}
+
{(\sum_{{\boldsymbol e}_j} \Delta^{-}_{i,{\boldsymbol e}_j} a_{i,{\boldsymbol e}_j})\Delta^2 u_i}
),
\end{eqnarray}
\begin{eqnarray}
\delta_B(\sum_i\frac{A}{2}(\partial_\tau\lambda_i)^2)&=&\sum_iA(\partial_\tau \lambda_i)\partial_\tau(\delta_B\lambda_i)=\epsilon\sum_i{A(\partial_\tau \lambda_i)(\partial_\tau^2 u_i)},
\\
\delta_B \sum_i (\frac{D}{2 }\sum_{{\boldsymbol e}_j}(\partial_\tau a_{i,i+{\boldsymbol e}_j})^2) 
&=&D\sum_i(\sum_{{\boldsymbol e}_j}\partial _\tau a_{i,i+{\boldsymbol e}_j}\partial _\tau \delta_B a_{i,i+{\boldsymbol e}_j})=\epsilon D\sum_i  {(\sum_{{\boldsymbol e}_j}\partial _\tau a_{i,i+{\boldsymbol e}_j}\partial _\tau (u_{i+{\boldsymbol e}_j}-u_i))}\nonumber\\
&=&\epsilon D\sum_i{\sum_{{\boldsymbol e}_j}(a_{i,i+{\boldsymbol e}_j}-a_{i-{\boldsymbol e}_j,i})\partial_\tau^2 u_i}=\epsilon D\sum_i{\sum_{{\boldsymbol e}_j} \Delta_{i,{\boldsymbol e}_j}^-a_{i,i+{\boldsymbol e}_j}\partial_\tau^2 u_i},
\\
\delta_B \sum_i (\frac{E}{2}\sum_{{\boldsymbol e}_j}(\lambda_{i+{\boldsymbol e}_j}-\lambda_i)^2) &=& \sum_i(E\sum_{{\boldsymbol e}_j}(\lambda_{i+{\boldsymbol e}_j}-\lambda_i)(\delta_B\lambda_{i+{\boldsymbol e}_j}-\delta_B\lambda_i))\nonumber\\
&=&\epsilon\sum_i {(E\sum_{{\boldsymbol e}_j}(\lambda_{i+{\boldsymbol e}_j}-\lambda_i)\partial_\tau (u_{i+{\boldsymbol e}_j}- u_i))}=\epsilon\sum_i {E\partial_\tau \lambda_i\Delta^2 u_i},
\\
\delta_B(\sum_i\frac{C}{2 } \sum_{{\boldsymbol e}_j}( \Delta_{i,{\boldsymbol e}_j}^-a_{i,i+{\boldsymbol e}_j})^2)&=&C\sum_i \sum_{{\boldsymbol e}_j}( \Delta_{i,{\boldsymbol e}_j}^-a_{i,i+{\boldsymbol e}_j})\delta_B(a_{i,i+{\boldsymbol e}_j}-a_{i-{\boldsymbol e}_j,i}),\nonumber\\
&=&\epsilon C\sum_i \sum_{{\boldsymbol e}_j}( \Delta_{i,{\boldsymbol e}_j}^-a_{i,i+{\boldsymbol e}_j})((u_{i+{\boldsymbol e}_j}-u_i)-(u_i-u_{i-{\boldsymbol e}_j}))\nonumber\\
&=&\epsilon \sum_i {C\sum_{{\boldsymbol e}_j}( \Delta_{i,{\boldsymbol e}_j}^-a_{i,i+{\boldsymbol e}_j})\Delta^2 u_i},\\
\delta_B (\sum_i B\sum_{{\boldsymbol e}_j} (\partial_\tau a_{i,i+{\boldsymbol e}_j}(\lambda_{i+{\boldsymbol e}_j}-\lambda_i)))&=&\sum_i B \sum_{{\boldsymbol e}_j} ((\partial_\tau \delta_B a_{i,i+{\boldsymbol e}_j})(\lambda_{i+{\boldsymbol e}_j}-\lambda_i)+ (\partial_\tau a_{i,i+{\boldsymbol e}_j}\delta_B(\lambda_{i+{\boldsymbol e}_j}-\lambda_i)))\nonumber\\
&=&\epsilon\sum_i( {B\partial_\tau \lambda_i\Delta^2 u_i} +{B\sum_{{\boldsymbol e}_j} \Delta_{i,{\boldsymbol e}_j}^-a_{i,i+{\boldsymbol e}_j}\partial_\tau^2 u_i}),
\end{eqnarray}
{up to some surface terms. To cancel these terms with the ghost part, the following condition should be satisfied,}
\begin{equation}
     {A=C\zeta^2,\quad C\zeta=B+D=B+E,\quad E=D},
\end{equation}
and the BRST symmetry is preserved on the lattice.

The equations of motion for $\lambda_i$, $a_{i,i+{\boldsymbol e}_j}$, and $u_i$ are
\begin{eqnarray}
\lambda: \quad&&A \partial^2_\tau \lambda_i+B \sum_{{\boldsymbol e}_j} \partial_\tau \Delta^+_{i,{\boldsymbol e}_j}a_{i,i+{\boldsymbol e}_j}+E\Delta^2 \lambda_i=
{\rm i}(\sum_\sigma f^\dag_{i\sigma}f_{i\sigma}+b^*_ib_i-1).\\
a_{i,i+{\boldsymbol e}_j}:\quad&&D\partial_\tau^2a_{i,i+{\boldsymbol e}_j}+C \Delta^2a_{i,i+{\boldsymbol e}_j}+B\partial_\tau \Delta^+_{i,{\boldsymbol e}_j}\lambda_i =J_{i,i+{\boldsymbol e}_j},\\
\bar{u}_i:\quad &&({\zeta}\partial_\tau^2+\Delta^2)u_i=0.
\end{eqnarray}
For the Noether current,
\begin{equation}
    \delta_B S=\sum_i\int {\rm d}\tau \partial_\mu(\frac{\partial \mathcal{L}}{\partial \partial_\mu \Phi_i}\delta_B  \Phi_i)=\epsilon\sum_i\int {\rm d}\tau \partial_\mu K^\mu.
\end{equation}
The BRST charge:
\begin{eqnarray}
   Q&=&\sum_i (\frac{\partial \mathcal{L}}{\partial \partial_\tau \Phi_i}\delta_B  \Phi_i-K^\tau)\nonumber\\
   &=&\sum_i(-{\rm i}g(\sum_\sigma f_{i\sigma}^\dag f_{i\sigma}+b^*_ib_i)u_i+(A\partial_\tau \lambda_i+(B+D)\sum_{{\boldsymbol e}_j} \Delta^-_{{\boldsymbol e}_j}a_{i,i+{\boldsymbol e}_j} )\partial_\tau u_i 
   +(-D\sum_{{\boldsymbol e}_j}\partial_\tau \Delta^-_{{\boldsymbol e}_j} a_{i,i+{\boldsymbol e}_j}+E\Delta^2 \lambda)u_i ).\nonumber\\
\end{eqnarray}
 {In the $E,D\rightarrow 0$ limit, the BRST charge Q and the current become
\begin{eqnarray}
    Q|_{E,D\rightarrow0}&=&\sum_i(-{\rm i}g(\sum_\sigma f_{i\sigma}^\dagger f_{i\sigma}+b^*_ib_i)u_i+(A\partial_\tau \lambda_i+B\sum_{{\boldsymbol e}_j} \Delta^-_{{\boldsymbol e}_j}a_{i,i+{\boldsymbol e}_j} )\partial_\tau u_i 
   ),\\
  J_{i,i+{\boldsymbol e}_j}|_{E,D\rightarrow 0}&=& C \Delta^2a_{i,i+{\boldsymbol e}_j}+B\partial_\tau \Delta^+_{i,{\boldsymbol e}_j}\lambda_i ,
\end{eqnarray}
which are the same as the direct lattice version  of the continuum limit (see Eqs. (\ref{Qcon}) and (\ref{Jcon}) in the main text) and induce the lattice versions of Gauss and current constraints.}

\section{An example of $\det ((D^{RPA}_{\mu\nu})^{-1})\neq 0$ in the limit of $D=E\rightarrow0$} \label{app2}

 {In the main text we argue that the skew $U(1)$ symmetry is broken in the presence of the matter field. To construct an explicit example of $D_{\mu\nu}$, for simplicity, we consider the gauge field is coupled to a relativistic Dirac field. Then due to the gauge symmetry, the vacuum polarization takes the following form,
\begin{equation}
    \Pi_{\mu\nu}=\Pi_{k^2}(k^2 g_{\mu\nu}-k_\mu k_\nu),
\end{equation}
where $\Pi_{k^2}$ is a function of momentum $k^2$.
   For  the gauge field part,
   \begin{equation}
       L_{GF}=\frac{A}{2}(\partial_\tau\delta\lambda)^2+B\sum_b \partial_\tau \delta a_b\partial_b\delta \lambda
		+\frac{C}{2}(\sum_b \partial_b \delta a_b)^2
		+\frac{D}{2}\sum_b (\partial_\tau \delta a_b)^2+\frac{E}{2}\sum_b(\partial_b \delta\lambda)^2,
   \end{equation}
   where $\tau=-{\rm i}t$. By substituting $({\rm i}\partial_t,{\rm i}\partial_{\boldsymbol r})=(\omega,-{\boldsymbol k})$, and choosing the metric $g_{\mu\nu}=(1,-1,-1,-1)$, 
the corresponding propagator $D^{(0)}_{\mu\nu}$ satisfies
\begin{eqnarray}
    (D^{(0)}_{\mu\nu})^{-1}&=&
    \begin{bmatrix}
        -\frac{A}{2}\omega^2+\frac{E}{2}(k_1^2+k_2^2) & -{\rm i}\frac{B}{2}\omega k_1 & -{\rm i}\frac{B}{2}\omega k_2 \\
       -{\rm i}\frac{B}{2}\omega k_1 & \frac{C}{2}k_1^2-\frac{D}{2}\omega^2 & \frac{C}{2}k_1 k_2 \\
       -{\rm i}\frac{B}{2}\omega k_2 & \frac{C}{2}k_1k_2 & \frac{C}{2}k_2^2-\frac{D}{2}\omega^2
    \end{bmatrix},\\
    \det( (D^{(0)}_{\mu\nu})^{-1})&=&-\frac{1}{8}D\omega^2((A\omega^2(D\omega^2-C (k_1^2+k_2^2)))+(k_1^2+k_2^2)(B^2\omega^2-DE\omega^2+CE(k_1^2+k_2^2))).
\end{eqnarray}
Here $\det( (D^{(0)}_{\mu\nu})^{-1})=0$ in the limit $D=E\rightarrow0$. As discussed in the main text, after coupling the gauge field to the matter field,  the inverse of the full propagator reads
\begin{eqnarray}
     (D'_{\mu\nu})^{-1}&=&(D_{\mu\nu}^{(0)})^{-1}-\Pi_{\mu\nu},\\
\det((D'_{\mu\nu})^{-1})&=&-\frac{1}{8}(D\omega^2+2\Pi_{k^2}(-\omega^2+k_1^2+k_2^2))((k_1^2+k_2^2)(\omega^2(B^2-4{\rm i}B\Pi_{k^2}+2(E-D)\Pi_{k^2}-DE)\nonumber\\
&&+(CE+2C\Pi_{k^2})(k_1^2+k_2^2))+A\omega^2((D-2\Pi_{k^2})\omega^2-C(k_1^2+k_2^2))).
\end{eqnarray}
 Denote $(D_{\mu\nu})^{-1}$ and $\Pi_{k^2}^*$ for the limit $D=E\rightarrow0$,
 \begin{eqnarray}
     \det ((D_{\mu\nu})^{-1})&=&-\frac{1}{4}\Pi_{k^2}^*(\omega^2-k_1^2-k_2^2)(A\omega^2(2\omega^2\Pi^*_{k^2}+C(k_1^2+k_2^2))\nonumber\\
    && -(k_1^2+k_2^2)(B^2\omega^2-4{\rm i}B\omega^2\Pi_{k^2}^*+2C\Pi_{k^2}^*(k_1^2+k_2^2))).
 \end{eqnarray}
By choosing RPA, $\Pi_{k^2}^*\rightarrow \Pi^{(0)*}_{k^2}$ which does not contain the contribution from the gauge field propagator, and $\det ((D^{RPA}_{\mu\nu})^{-1})\neq 0$.}

\end{widetext}

\end{document}